\documentclass[twocolumn,tighten]{aastex62}
\usepackage{amssymb, amsmath,framed}

\usepackage{bm}
\expandafter\ifx\csname package@font\endcsname\relax\else
 \expandafter\expandafter
 \expandafter\usepackage
 \expandafter\expandafter
 \expandafter{\csname package@font\endcsname}
\fi
\def\bq{\begin{equation}}
\def\eq{\end{equation}}
\def\bqy{\begin{eqnarray}}
\def\eqy{\end{eqnarray}}

\begin{document}
\title{\large{Revisiting the Biological Ramifications of Variations in Earth's Magnetic Field}}

\correspondingauthor{Manasvi Lingam}
\email{manasvi.lingam@cfa.harvard.edu}

\author{Manasvi Lingam}
\affiliation{Institute for Theory and Computation, Harvard University, Cambridge MA 02138, USA}

\begin{abstract}
An Earth-like planetary magnetic field has been widely invoked as a requirement for habitability as it purportedly mitigates the fluxes of ionizing radiation reaching the surface and the escape of neutrals and ions from the atmosphere. Recent paleomagnetic evidence indicates that the nucleation of Earth's inner core, followed perhaps by an increase in geomagnetic field strength, might have occurred close to the Ediacaran period. Motivated by this putative discovery, we explore the ensuing ramifications from the growth or reversals of Earth's dynamo. By reviewing and synthesizing emerging quantitative models, it is proposed that neither the biological radiation dose rates nor the atmospheric escape rates would vary by more than a factor of $\sim 2$ under these circumstances. Hence, we suggest that hypotheses seeking to explain the Cambrian radiation or mass extinctions via changes in Earth's magnetic field intensity are potentially unlikely. We also briefly discuss how variations in the planetary magnetic field may have impacted early Mars and could influence exoplanets orbiting M-dwarfs.\\
\end{abstract}

\section{Introduction} \label{SecIntro}
The modality and timing of the Earth's dynamo (geodynamo) remains the subject of intensive research \citep{McP98,TBM14}. In particular, much attention has been devoted to understanding when the inner core formed (nucleation). The reason is that this process, by virtue of releasing latent heat during crystallization and causing chemical differentiation, can provide the requisite power for driving the geodynamo. Some recent analyses favor inner-core nucleation during the Mesoproterozoic ($\sim 1$-$1.5$ Ga) based on paleomagnetic intensity measurements \citep{BPP15}, whereas other favor a Neoproterozoic origin \citep{Dris16}. More recently, \citet{BTN19} analyzed plagioclase and clinopyroxene crystals from the Sept {\^I}les Intrusive Suite from the Ediacaran period ($\sim 565$ Ma) and found that the Earth's magnetic moment was less than $10\%$ of its modern value. Based on $14$ data sets, \citet{BTN19} argued that their findings: (i) are consistent with geodynamo simulations yielding inner core crystallization during the Ediacaran, and (ii) might mark the initiation of a geodynamo with $\sim 200$ Myr cycles.

If the geomagnetic field strength increased during the Cambrian period, it may explain the famous Cambrian evolutionary radiation \citep{Mar06}. Most studies that propose this hypothesis bank on two different, although interconnected, effects ostensibly engendered by the development of a strong magnetic field \citep{DPC16,MLB16}. The first is added protection against atmospheric escape that permits the retention of a thicker oxygenated atmosphere and diminished doses of ultraviolet (UV) radiation reaching the surface. The second is the deflection of Galactic Cosmic Rays (GCRs) and other energetic particles, thereby reducing the flux of ionizing radiation penetrating to the surface. These dual reasons serve to explain why strong magnetic fields are widely considered an important (perhaps even necessary) component of planetary habitability \citep{Cock16,LL19}.

In this Letter, we will therefore explore the potential ramifications that may arise if the Earth had transitioned from a weakly magnetized regime during the Ediacaran to the current geodynamo. More broadly, we revisit the consequences for habitability that would result from turning the magnetic moment ``on'' or ``off'' in the context of surface radiation dose rates and atmospheric escape for Earth, Mars and exoplanets.

\section{Magnetospheric Shielding and Radiation Dose Rates}\label{SecGCR}
We will examine how magnetospheric shielding, and the radiation dose rate at the surface originating from GCRs, could be affected by the onset of the geodynamo. The magnetopause distance scales as
\begin{equation}\label{Rmpd}
    R_{mp} \propto \left(\frac{\mathcal{M}^2}{ P_{sw}}\right)^{1/6},
\end{equation}
where $\mathcal{M}$ represents the magnetic moment of the planet, while $P_{sw}$ denotes the solar wind pressure at this location \citep{Gom98}. The Earth's magnetopause changes due to fluctuations in $P_{sw}$ \citep{SLR91}, but we may roughly express it as follows:
\begin{equation}
    R_{mp} \approx 10.7\,R_\oplus\,\left(\frac{\mathcal{M}(t_\star)}{\mathcal{M}_0}\right)^{1/3} \left(\frac{P_{sw}(t_\star)}{P_0}\right)^{-1/6},
\end{equation}
where $\mathcal{M}_0 \approx 7.9 \times 10^{22}$ A m$^2$ denotes the current magnetic moment of the Earth and $P_0 \approx 3.2 \times 10^{-9}$ Pa is the modern solar wind pressure at the vicinity of the Earth \citep{SN09}. Next, we observe that $P_{sw}$ is apparently dominated by the dynamical pressure at all epochs \citep{BLK10,DHL17}. Hence, it is reasonable to assume that $P_{sw} \propto n_{sw} v_{sw}^2$, where $n_{sw}$ and $v_{sw}$ represent the number density and velocity of the solar wind close to the Earth. We adopt the theoretical scalings $n_{SW} \propto t_\star^{-1.54}$ and $v_{sw} \propto t_\star^{-0.4}$ \citep{BLK10}, with $t_\star$ denoting the stellar age. Thus, with these simplifications, we obtain
\begin{equation}\label{Rmptvar}
    R_{mp} \approx 10.7\,R_\oplus\,\left(\frac{\mathcal{M}(t_\star)}{\mathcal{M}_0}\right)^{1/3} \left(\frac{t_\star}{t_0}\right)^{0.39},
\end{equation}
where $t_0 \approx 4.6$ Gyr is the Sun's current age. The above scaling is consistent with the theoretical model developed in \citet{TCW10} that was based on a different, albeit related, approach. If we adopt the recent proposal that $\mathcal{M}/\mathcal{M}_0 \approx 0.09$ \citep{BTN19} in the Ediacaran (corresponding to $t_\star \approx 4.0$ Gyr) and substitute them into (\ref{Rmptvar}), we obtain $R_{mp} \approx 4.6\,R_\oplus$; in other words, the magnetopause distance would be compressed to approximately $43\%$ of its present-day value. This value closely agrees with the prediction that the magnetopause distance was $\lesssim 4.5\,R_\oplus$ \citep{BTN19}.

Given the decreased magnetospheric shielding, we can ask ourselves how the accompanying dose of ionizing radiation reaching the surface is impacted. Before doing so, it is essential to recognize that any magnetized planet offers two distinct layers of shielding above its surface: the magnetosphere and the atmosphere. Therefore, before investigating the effects of a diminished magnetic moment, the atmospheric shielding merits consideration. The atmospheric column density ($\Sigma$) equals
\begin{equation}\label{ColDens}
    \Sigma = \frac{P_s}{g},
\end{equation}
where $P_s$ denotes the surface pressure. As $g$ evidently remains unaffected for a given planet, we have $\Sigma \propto P_s$. There are many uncertainties surrounding the composition of the Ediacaran atmosphere, most notably the partial pressure of molecular oxygen \citep{LRP14,CK17}. Nonetheless, multiple lines of evidence seemingly indicate that the total atmospheric pressure was probably close to that of the modern value. \citet{OSRL18} recently synthesized constraints from numerical models and geochemical proxies (e.g., N isotopes, basalt vesicles) to estimate the total atmospheric pressure over geological time. From Figure 4 of \citet{OSRL18}, the total atmospheric pressure was $0.9$-$1.2$ bar in the Phanerozoic ($< 0.54$ Ga) whereas it appears to have ranged between $\sim 0.5$-$1.5$ bar during the ``Boring Billion'' ($0.8$-$1.8$ Ga). Hence, based on these putative limits, the surface pressure deviated by less than a factor of $2$ from the canonical value of $1$ bar. Consequently, invoking (\ref{ColDens}), it is reasonable to suppose that $\Sigma$ in the Ediacaran was close to its value today.

Thus, we can now turn our attention to the effects arising from lowering the magnetic moment of the Earth, while holding $\Sigma$ roughly constant. GCRs induce the production of secondary particles such as muons, pions and electrons in the atmosphere. The resultant ionizing radiation that reaches the surface causes damage to biomolecules as well as to organisms \citep{Dart11}. Therefore, due to the smaller magnetic moment, higher biological dose rates ought to follow. \citet{GTS16} carried out numerical simulations to estimate the fluxes of secondary particles and the ensuing dose rates as a function of the magnetic moment and column density.\footnote{A potentially important caveat is that \citet{GTS16} studied Earth-like planets orbiting a $0.45\,M_\odot$ star that would have plausibly experienced a more intense stellar wind.} In particular, for Earth-like atmospheric column densities, the biological radiation dose rate ($\mathcal{B}$) was increased by a factor of $2$ for a completely unmagnetized planet and by $40\%$ for $\mathcal{M} \approx 0.1\,\mathcal{M}_\odot$. From the data in Table 2 of \citet{GTS16} for an Earth-like atmosphere, we adopt the following ansatz after using the NonlinearModelFit routine from {\scriptsize {MATHEMATICA}}:
\begin{equation}\label{BRDR}
   \frac{\mathcal{B}}{\mathcal{B}_0} \approx \left[1 + 0.95 \left(\frac{\mathcal{M}}{\mathcal{M}_0}\right)^{0.56}\right]^{-1}, 
\end{equation}
where $\mathcal{B}_0 \approx 6.5 \times 10^{-4}$ Sv/yr is the biological radiation dose rate for a completely unmagnetized planet. 

As per the prior discussion and results, the elevated dose rates are likely to have minimal effects on biota. Astrophysical phenomena, such as solar proton events (SPEs), could drive greater increases in fluxes of ionizing radiation and UV radiation reaching the surface \citep{GTS16} compared to decreasing the magnetic moment. We will use (\ref{BRDR}) to deduce the consequences for mutations. The most commonly employed model in the literature to describe the mutagenic response to ionizing radiation is the linear response model based on theoretical arguments as well as experimental data \citep{Thack92,NL99,BDG03}. Broadly construed, this model suggests that the mutation rate might be linearly proportional to the radiation dose rate. However, because the background radiation dose is typically much lower than those utilized in lab experiments, the mutation rates may not follow a linear trend \citep{HBD04}.

Bearing this caveat in mind, the linear model in conjunction with (\ref{BRDR}) implies that mutation rates will only be elevated by $50\%$-$100\%$ if the Earth's magnetic moment is reduced to roughly $10\%$ of its value. To reverse the premise, if the Earth had transitioned from  $0.1\,\mathcal{M}_0$ to $\mathcal{M}_0$, the corresponding decrease in mutation rates would only be a factor of $1.5$-$2$. We can thus ask ourselves if this decrease has any tangible effects on today's biota. For resolving this question, we turn to the famous ``error threshold'' introduced in the seminal work by \citet{Eig71}. The central idea is that the product of the mutation rate per base ($\mathcal{U}$) for a particular species and the length of that organism's genome ($L$) ought to be generally smaller than unity in order to enable adaption \citep{ES79}; exceeding this threshold will lead to breakdown in evolutionary optimization. Although the error threshold is a theoretical construct and does not apply to all viruses, it has been empirically validated for a large number of current species on Earth \citep{BE05,Now06}. For a given mutation rate, the critical length $L_c$ for the onset of this threshold is therefore given by
\begin{equation}\label{Lcrit}
    L_c = \frac{1}{\mathcal{U}}.
\end{equation}

To evaluate the consequences of a reduced magnetic moment, it is instructive to carry out the following thought experiment. If we take modern organisms and place them in an environment with higher ambient mutation rates, we can determine whether the error threshold would be exceeded. The product $\Delta = \mathcal{U} L$ for \emph{Escherichia coli}, \emph{Saccharomyces cerevisiae} (yeast), \emph{Drosophila melanogaster}, mice and humans is $2.5 \times 10^{-3}$, $2.7 \times 10^{-3}$, $5.8 \times 10^{-2}$, $0.49$ and $0.16$, respectively \citep{Now06}. Raising $\mathcal{U}$ approximately by a factor of $2$ as per the preceding paragraph still ensures that $\Delta < 1$ is preserved for all the above species.\footnote{Recall that specifying this boost implicitly amounts to postulating that $\mathcal{U}$ exhibits a roughly linear dependence on the ambient biological radiation dose rate.} Likewise, we can make use of (\ref{Lcrit}) to determine the maximum genome length feasible for a given mutation rate. If $\mathcal{U}$ is proportional to $\mathcal{B}$, from (\ref{BRDR}) we see that moving from $0.1\,\mathcal{M}_0$ to $\mathcal{M}_0$ merely yields an increase in $L_c$ by $\sim 60\%$ compared to what its value would be today. 

In actuality, doubling the mutation rate may represent an overestimate because GCRs are not the only continuous source of ionizing radiation; radiogenic decay of elements in the Earth's crust contributes to a biological radiation dose rate of $2.4 \times 10^{-3}$ Sv/yr \citep{HSW09}. Furthermore, aquatic organisms are shielded against cosmic rays by water; the CR flux is further reduced by a factor of $\gtrsim 2$-$3$ after passing through $\sim 1$ m of water \citep{AKA11}. Hence, viewed collectively, reducing the Earth's magnetic field is not expected to significant impact present-day species. On the other hand, in the unlikely event that all complex multicellular biota functioned very close to the error threshold, the modest decrease in mutation rates which would arise if the geodynamo was amplified at this stage, might have facilitated the Cambrian radiation.

\section{Atmospheric escape rates}\label{SecAER}
We explore how atmospheric escape rates would be altered by the strength of the planet's magnetic field. In case the atmospheric escape rate is greatly enhanced, this should lead to two major consequences. First, attenuated atmospheres can permit higher fluxes of GCRs, thus resulting in greater radiation doses. Second, the flux of UV radiation reaching the surface may increase for thinner atmospheres, although this will also depend on atmospheric composition.

A simple model for the atmospheric escape rate ($\dot{M}$) for \emph{unmagnetized} planets was proposed by \citet{ZSR10} based on the conservation of mass flux:
\begin{equation}\label{MLRate}
\dot{M}  \approx \frac{1}{4}\left(\frac{R}{a}\right)^2 \dot{M}_\star,
\end{equation}
where $R$ and $a$ are the planet's size and orbital radius, respectively. Note that $\dot{M}_\star$ denotes the time-dependent stellar mass loss rate. This model is in good agreement with the total escape rate for modern Mars deduced from MAVEN data and multi-fluid magnetohydrodynamic (MHD) simulations \citep{LL19}. Moreover, it captures the general trend of how atmospheric escape changes with $R$, $a$ and $t_\star$ reasonably well \citep{DJL18,DLM18}. Using the stellar mass loss scaling $\dot{M}_\star \propto t_\star^{-2.33}$ from Ly$\alpha$ absorption \citep{WM05}, we have
\begin{equation}\label{MLRatefin}
\frac{\dot{M}}{\dot{M}_0}  \approx \frac{1}{4}\left(\frac{R}{a}\right)^2 \left(\frac{t_\star}{t_0}\right)^{-2.33},
\end{equation}
where $\dot{M}_0$ represents the current atmospheric escape for an unmagnetized planet. Substituting $t_\star \approx 4$ Gyr and $t_0 \approx 4.6$ Gyr, we obtain $\dot{M}/\dot{M}_0 \approx 1.38$. This calculation helps build an intuition for the evolution of atmospheric escape from an unmagnetized planet over time. 

What we seek to study, however, are atmospheric escape rates for magnetized planets. It is difficult to resolve this matter definitively as numerous physical mechanisms are responsible for the escape of neutrals and ions. Some notable processes include Jeans escape, sputtering, ion pickup, escape from polar caps and cusps, and dissociative recombination \citep{Lam13}. Consequently, there has been a dawning realization that decreasing the magnetic moment may actually translate to an \emph{decrease} in the atmospheric escape rates, \emph{contra} previous expectations. We shall illustrate our point by focusing on two recent studies that have investigated atmospheric escape from magnetized planets.

\citet{BT18} calculated the upper bounds on atmospheric loss for Earth-like planets in Section 4.2 of their work. The final expression is complicated as it not only depends on $R$, $a$ and $R_{mp}$ and $v_{sw}$ (that were introduced earlier), but also on: (i) incoming sound speed $c_s$, (ii) speed of magnetic reconnection on the wind-side, and (iii) exobase radius. We hold (i)-(iii) roughly constant along the lines of \citet{BT18} and adopt Equation (26) from that paper. Thus, by doing so, we obtain
\begin{equation}
    \dot{M} \propto \dot{M}_\star v_{sw} R_{mp}^2.
\end{equation}
As before, we normalize the variables in this equation by their present-day values, thus ending up with
\begin{equation}\label{MagPRat}
 \frac{\dot{M}}{\dot{M}_0}  \approx \left(\frac{\mathcal{M}(t_\star)}{\mathcal{M}_0}\right)^{2/3} \left(\frac{t_\star}{t_0}\right)^{-1.95},
\end{equation}
where we have made use of (\ref{Rmptvar}) along with the scaling relations for $\dot{M}_\star$ and $v_{sw}$. If we make use of $\mathcal{M} \approx 0.09\,\mathcal{M}_0$ as well as $t_\star \approx 4$ Gyr and $t_0 \approx 4.6$ Gyr, we find $\dot{M}/\dot{M}_0 \approx 0.26$. Because $\dot{M}_0$ represents the modern atmospheric escape rate, this result suggests that the maximal loss rates might have actually been reduced at $\sim 0.6$ Ga due to the lower magnetic moment. 

Note that the enhancement in escape rates due to the dynamical evolution of stellar wind parameters, as exemplified by the last factor on the right-hand side of (\ref{MLRatefin}) and (\ref{MagPRat}), is very close to unity if we consider the Ediacaran period. Hence, we can set aside this factor to focus on understanding how atmospheric escape changes with the magnetic moment. In a recent theoretical study, \citet{GMN18} considered a diverse array of thermal and non-thermal escape processes to determine the escape rate as a function of the magnetic moment. From Figure 2 of \citet{GMN18}, we observe that $\dot{M} \approx 1.6$ kg/s at $\mathcal{M} \approx 0.1\,\mathcal{M}_0$ whereas $\dot{M}_\oplus \approx 1.4$ kg/s at $\mathcal{M} = \mathcal{M}_0$; interestingly, for $\mathcal{M} < 0.01 \mathcal{M}_0$, we notice that the escape rate declines to $\dot{M} \approx 1.2$ kg/s.

Hence, as per this model, acquiring a stronger magnetic field may reduce escape rates by $\mathcal{O}(10\%)$, but clearly this difference is minimal. In fact, the maximal atmospheric escape rate of $3.4$ kg/s at $\mathcal{M} \approx 0.01\,\mathcal{M}_0$ is only a factor of $2.4$ higher than the escape rate at $\mathcal{M} = \mathcal{M}_0$. In the regime $0.01 \lesssim \mathcal{M}/\mathcal{M}_0 < 1$, the escape from polar caps dominates. We can therefore make use of the empirical relation
\begin{equation}
  \frac{\dot{M}}{\dot{M}_\oplus}  \approx 19.3 \left[1 - \sqrt{1 - 0.1\left(\frac{\mathcal{M}}{\mathcal{M}_0}\right)^{-1/3}}\right],  
\end{equation}
based on Figure 2 and Equation (A.5) of \citet{GMN18}. This formula captures the trend quite accurately in the aforementioned regime and overestimates the peak merely by a factor of $2$. 

Thus, based on theoretical models, there are grounds for supposing that transitioning from a weaker to stronger magnetic field did not impact the atmospheric escape rates, and therefore the surface pressure, by a significant margin. In this context, note that geochemical proxies indicate that the surface pressure might have changed by less than a factor of $\sim 2$ since the mid-Proterozoic; see Section \ref{SecGCR} for further details. 

\section{Discussion}
Here, we briefly discuss other implications of varying magnetic fields for biota on Earth and other worlds. For the sake of clarity, we shall suppose the transition from $\mathcal{M} \lesssim 0.1\,\mathcal{M}_0$ to $\mathcal{M} = \mathcal{M}_0$, or vice-versa, happens rapidly enough to ensure that stellar wind parameters remain mostly unaffected.

\subsection{Solar Cosmic Rays (SCRs)}
Hitherto, we have focused only on GCRs and ignored the contribution of energetic particles, henceforth known as SCRs, generated during SPEs. The biologically weighted UV flux at the surface can increase by an order of magnitude compared to the background during SPEs \citep{GTS16,TGG16}. It should, however, be recognized that these events are sporadic, unlike GCRs that impact the Earth continuously. Furthermore, SCRs are typically in the keV-MeV range, reaching a maximum of $\sim 10$ GeV, and therefore blocked more easily either by Earth-like atmospheres or magnetic fields. The time-averaged energy and number fluxes of SCRs were estimated to be smaller than the corresponding values for GCRs by up to one order of magnitude even for the active young Sun at $\sim 4$ Ga \citep{LDF18}.

Let us further examine the case wherein all parameters of the Earth barring the magnetic moment are held constant. For the 1989 SPE event, \citet{Atri} found that the biological radiation dose decreased by only a factor of $\sim 3$ as one moved from $\mathcal{M} \approx 0.1\,\mathcal{M}_0$ to $\mathcal{M} \approx \mathcal{M}_0$. It underscores the point that the change in biological radiation dose due to SCRs, which would result from modifying the magnetic moment, is not significant. Moreover, the radiation dose was only between $10^{-6}$ and $10^{-5}$ Sv across this range, whereas the critical value for humans is $\sim 5$-$10$ Sv \citep{Atri}. SPEs associated with solar superflares whose energies are $\gtrsim 10^{29}$ J may adversely affect ecosystems, but their frequency and probability remains indeterminate \citep{LL17b}.

\subsection{Geomagnetic reversals}
We have addressed the transition from $\mathcal{M} \lesssim 0.1\,\mathcal{M}_0$ to $\mathcal{M} = \mathcal{M}_0$ specifically with regards to Ediacaran paleomagnetic data. Note, however, that this transition could also apply to certain scenarios of geomagnetic reversals \citep{McP98}. In principle, the magnetic field can approach values close to zero during the reversal process. For example, sediment cores from the Black Sea imply that the magnetic field during the Laschamp geomagnetic polarity excursion may have dropped to $5\%$ of its maximal value \citep{NAF12}. Therefore, several analyses have proposed that geomagnetic reversals are inextricably linked to mass extinctions, either via increased flux of ionizing radiation \citep{Uff63} or rapid escape of oxygen \citep{WPZ14} due to the absence of protection accorded by the magnetic field. 

While much work has been done to identify potential correlations between mass extinctions and geomagnetic reversals, the evidence for a causal link between the latter and former remains elusive, perhaps even non-existent \citep{GV10}. Our preceding analysis showed that changes in biological radiation dose rate and the atmospheric escape rate were probably minimal even during the transition from a completely unmagnetized Earth to one endowed with its current magnetic moment. Hence, our results appear to be consistent with the absence of unambiguous causal connections between geomagnetic field reversals and mass extinctions, because we predict that the ramifications arising from the former are not sufficiently impactful.

\subsection{Mars and Exoplanets around M-dwarfs}
Because we have made the case that changing the magnetic moment does not significantly alter Earth's ecosystems, it may seem natural to presume that the same would hold true for other planets. However, there is one subtle point worth reiterating: the atmospheric column density during the transition (e.g., dynamo onset or reversals) is crucial.

For example, consider a roughly Earth-sized planet with a surface pressure of $0.1$ bar. Its column density would therefore be $\sim 10\%$ of current Earth as per (\ref{ColDens}). For such a world, the biological radiation dose rate scales differently with $\mathcal{M}$ as compared to (\ref{BRDR}). From Table 2 of \citet{GTS16}, after employing {\scriptsize {MATHEMATICA}}, we construct the ansatz:
\begin{equation}\label{BRDRv2}
   \frac{\mathcal{B}}{\mathcal{B}'_0} \approx \left[1 + 2.1 \left(\frac{\mathcal{M}}{\mathcal{M}_0}\right)^{2}\right]^{-1}, 
\end{equation}
where $\mathcal{B}'_0 \approx 0.553$ Sv/yr is the radiation dose rate for $\mathcal{M} \approx 0$. Therefore, we see that the dose rate changes by a factor of more than $3$ as one moves from $\mathcal{M} = 0$ to $\mathcal{M} = \mathcal{M}_0$; this change is $1.6$ times higher than what we obtained in Section \ref{SecGCR} for Earth-like column densities.

Let us turn our attention to Mars. It is well known that modern Mars does not have a global dipole field; it exhibits remnant crustal magnetic fields instead \citep{ACN99}. The time at which the Martian dynamo stopped is subject to uncertainty with some earlier studies indicating that this event took place $< 4$ Ga \citep{SRM00}. Recent analyses of the magnetization states of impact craters at different ages indicate that the shutdown occurred at $4.0$-$4.1$ Ga \citep{Lil13}. During this era, geological proxies are seemingly consistent with, but do not necessarily guarantee, an atmosphere of $\sim 1$ bar \citep{RC18,kite19}. Therefore, if Mars underwent a transformation from $\mathcal{M} \sim \mathcal{M}_0$ to $\mathcal{M} \sim 0$, the biological dose rate would be enhanced merely by $\mathcal{O}(1)$ based on prior discussion.

Next, we must consider the impact on atmospheric escape rates. The analysis of argon isotope ratios, reflecting escape due to ion pickup, has revealed that $> 66\%$ of atmospheric argon has been lost to space \citep{Jak17}, but it is not easy to pinpoint when and how the Martian atmosphere was depleted \citep{EAA}. In Section \ref{SecAER}, we saw that the reduction in the magnetic moment does not automatically translate to much higher escape rates. Multi-fluid MHD simulations have recently illustrated that the escape rate of ions from the Martian atmosphere is a non-monotonic function of the magnetic field, implying that atmospheric losses were potentially higher when Mars possessed a strong dipole field \citep{DLML18}; in fact, Martian atmospheric escape could be enhanced even for weak magnetic fields \citep{SS18}. Thus, the shutdown of the Martian dynamo might not have been as detrimental to short-term habitability as commonly supposed, although its long-term effects were clearly profound.

Lastly, a comment on temperate exoplanets around M-dwarfs is in order. Owing to their closer distances and stronger stellar winds from the host star, the atmospheric escape rates are probably a few orders of magnitude higher relative to Earth \citep{DLMC17,DJL18}. Hence, some exoplanets may manifest depleted atmospheres with low column densities. The shutdown of the planetary dynamo can therefore preferentially amplify biological dose rates relative to planets with Earth-like atmospheres. Moreover, biological dose rates in both the unmagnetized and magnetized cases are $2$-$3$ orders of magnitude higher than those received by Earth \citep{GTS16}. Note that M-dwarfs are also generally characterized by higher activity, thereby resulting in regular flares and SPEs that further elevate radiation doses at the surface \citep{LL19}. Hence, it is not unreasonable to surmise that the onset or shutdown of the planetary dynamo would have a comparatively higher impact on the habitability of such worlds with respect to Earth. 

\section{Conclusion}
Motivated by emerging evidence that favors the nucleation of the inner core during the Ediacaran, we investigated how putative variations in the magnetic moment may have affected Earth's habitability. 

First, we considered the possibility that increasing the magnetic field strength led to a substantial decrease in the surficial flux of ionizing radiation. We illustrated that the change in biological radiation dose rate was $\lesssim 2$ by utilizing recent numerical models. This prediction might be consistent with the cosmogenic isotope record (specifically $^{14}$C and $^{10}$Be) reconstructed from meteorites and lunar rocks, because the latter ostensibly implies that the average GCR flux has varied by a factor of $< 1.5$ over the past $\sim 1$ Gyr \citep{Uso17}. Next, by drawing upon the error threshold paradigm, we determined that most of the current species could have withstood higher radiation doses. Hence, unless the majority of complex multicellular biota were much more susceptible to DNA damage and mutagenesis, it seems unlikely that the slightly elevated radiation dose rates markedly influenced evolution. 

Second, we drew upon ongoing developments in atmospheric escape mechanisms to explore whether a sudden increase in the magnetic moment will lower escape rates. We found that atmospheric losses are perhaps reduced, but only by a factor of $< 2.5$ compared to the weakly magnetized case. Therefore, we do not anticipate the resultant variations to contribute substantially to atmospheric evolution. Subsequently, we discussed the impact of SCRs on biota and argued that their cumulative impact is usually lower than GCRs. We followed this up by indicating how our reasoning is applicable, after due modifications, to geomagnetic field reversals as they may be characterized by rapid changes in the magnetic moment. Finally, we discussed how the shutdown of the Martian dynamo influenced the habitability of early Mars (possibly causing marginal short-term impact) and the greater vulnerability of M-dwarf exoplanets to the initiation or shutdown of their dynamos.

In summary, our analysis suggests that the nucleation of the inner core or an increase in the magnetic field strength were probably not the drivers of the Cambrian radiation. Instead, it appears more plausible that the highly dynamic and heterogeneous Ediacaran geochemical environment, which emerged in the aftermath of the Marinoan glaciation, may have facilitated rapid evolutionary innovations, speciation and diversification \citep{Knoll15,DTG17,BJ17}.

\acknowledgments
The reviewer is thanked for providing insightful suggestions concerning this Letter. This work was supported by the Breakthrough Prize Foundation, Harvard University's Faculty of Arts and Sciences, and the Institute for Theory and Computation (ITC) at Harvard University. 


\end{document}